\documentclass[aps,prl,unsortedaddress,showpacs,floatfix,nofootinbib,twocolumn]{revtex4-1}
\usepackage[colorlinks=true,breaklinks=false]{hyperref}
\usepackage{graphicx}
\usepackage[normalem]{ulem}
\usepackage{amsmath,amstext,amssymb}
\usepackage{dcolumn}
\usepackage{bm}
\usepackage[english]{babel}
\usepackage{xcolor}

\newcommand{\mic}{~\mu {\rm m}}
\usepackage{gensymb}
\usepackage{multibib}

\begin{document}
\title{Octave-spanning supercontinuum generation in a silicon-rich nitride waveguide}

\author{Xing Liu$^1$, Minhao Pu$^1$, Binbin Zhou$^1$, Clemens J. Kr{\"u}ckel$^2$, Attila F{\"u}l{\"o}p$^2$, Victor Torres-Company$^{2,*}$, and Morten Bache$^{1,\dagger}$,}

\affiliation{$^1$DTU Fotonik, Department of Photonics Engineering, Technical University of Denmark, DK-2800 Kgs. Lyngby, Denmark\\
$^2$Department of Microtechnology and Nanoscience, Chalmers University of Technology, SE-41296 Gothenburg, Sweden\\
$^*$torresv@chalmers.se\\$\dagger$moba@fotonik.dtu.dk\\
\today}




\begin{abstract}

We {\color{black}experimentally show octave-spanning} supercontinuum generation in a non-stoichiometric silicon-rich nitride waveguide when pumped by femtosecond pulses from an erbium fiber laser. The pulse energy and bandwidth are comparable to results achieved in stoichiometric silicon nitride waveguides, but our material platform is simpler to manufacture. We also observe wave-breaking supercontinuum generation by using orthogonal pumping in the same waveguide. Additional analysis reveals that the waveguide height is a powerful tuning parameter for generating mid-infrared dispersive waves while keeping the pump in the telecom band.
\end{abstract}


\maketitle


\noindent
Supercontinuum generation in photonic crystal fibers 
{\color{black}
pumped with a mode-locked laser oscillator allowed generating octave-broadened coherent spectra without an amplifying stage \cite{Ranka2000}. This remarkable achievement
paved way for self-referencing in laser frequency combs \cite{Jones2000} and key advances in biophotonics \cite{Hartl2001}.}

There is a growing interest in supercontinuum generation in photonic integrated waveguides. This offers the prospect of supercontinuum generation in different regions of the electromagnetic spectrum with extremely low pump pulse energies delivered by near-IR ultrafast femtosecond fiber lasers.
One of the most promising waveguide platforms that has demonstrated near-IR-pumped octave-spanning {\color{black}supercontinua} is silicon nitride \cite{Halir2012,Zhao2015,Epping2015-SCG,Johnson2015,Mayer2015,Boggio2014}, 
{\color{black}in which} waveguides can be manufactured directly on an oxidized silicon wafer{\color{black}, and which, unlike} silicon, has negligible two-photon absorption in the near IR. The nonlinear Kerr coefficient is comparatively lower than other integrated waveguide technologies, but this is offset by lower linear propagation loss. As a result, large nonlinear phase shifts can be obtained {\color{black}at} modest pump powers. The broadest supercontinuum generated on chip has been measured on a silicon nitride
waveguide pumped with an Yb mode-locked fiber laser \cite{Epping2015-SCG}. Carefully engineered silicon nitride
waveguides {\color{black}provided} highly coherent supercontinua \cite{Johnson2015}, {\color{black}suitable for detecting the carrier envelope offset frequency of frequency combs \cite{Mayer2015}.}

Thick waveguides are needed to achieve suitable dispersion engineering and high mode confinement. However, silicon nitride films tend to crack above 400 nm thickness. Advanced fabrication methods utilize crack barriers and in-trench growth to overcome film cracking \cite{Luke2013,Epping2015} which recently enabled the supercontinua reported in  \cite{Epping2015-SCG,Johnson2015,Mayer2015}. An octave-spanning spectrum in non-stoichiometric silicon nitride waveguides was recently reported but it required a complex multi-cladding structure for dispersion engineering \cite{Boggio2014}. A simpler approach to circumvent cracking is to reduce film stress by slightly increasing the silicon content during deposition.  This allows thick crack-free film growth in a single deposition step with minimized process complexity. The resulting so-called silicon-rich nitride \cite{Philipp2004} shows a higher refractive index and an increased material absorption loss than stoichiometric silicon nitride, but this is compensated for by  a higher nonlinear Kerr coefficient
\cite{Wang2015,Krueckel2015a}.

Here we report octave-spanning supercontinuum generation in a 10 mm long
silicon-rich nitride waveguide \cite{Krueckel2015a} pumped in the quasi-TE mode with a mode-locked femtosecond Er-fiber laser in the telecom C band.
The key waveguide aspects are that an increased nonlinearity compensates for the higher propagation loss, and that thick waveguides can be made with a simple, single-cladding, CMOS-compatible structure suitable for dispersion engineering at telecom wavelengths. This gives a supercontinuum performance similar to stoichiometric silicon nitride waveguides made with advanced fabrication techniques \cite{Halir2012,Zhao2015,Epping2015-SCG,Mayer2015,Johnson2015}. When pumping in the quasi-TM mode, spectral broadening from optical {\color{black}wave breaking} is
observed for the first time in silicon nitride waveguides, {\color{black}which we show is caused 
by the TM pump mode experiencing normal dispersion.}

\begin{figure}[tbp]
\centering
\includegraphics[width=0.81\linewidth]{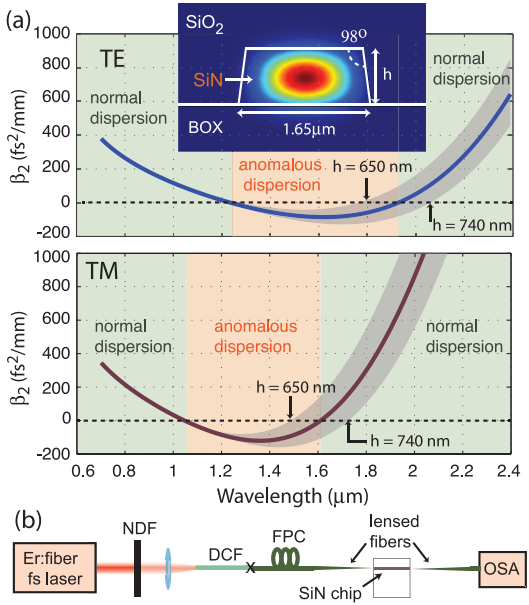}
\caption{(a) Numerically calculated GVD for the TE and TM modes for $h$ = 695 nm. The shadowed area indicates the variations from changing the waveguide height $h=650$-740 nm; this changes the second ZDW from 1.7 to 2.1 $\mic$ (see also Fig. \ref{fig:ZDW}).
(Inset) Cross-section geometry and modal confinement of power for fundamental quasi-TE mode. (b) Experimental setup for supercontinuum generation. NDF, neutral density filter; DCF, dispersion compensating fiber;  FPC, fiber polarization controller; OSA, optical spectrum analyzer.}
\label{fig:SiN}
\end{figure}

Our silicon-rich nitride film is fabricated in a low-pressure chemical vapor deposition (CVD) process \cite{Krueckel2015a}. The waveguide patterns are transferred by standard deep UV contact lithography followed by dry etching. The silica top cladding is done in a plasma-enhanced CVD process. The waveguides feature a Kerr parameter of
$\gamma=5.7
\pm 0.5~{\rm  (Wm)}^{-1}$
around $1.55 \mic$, which is a slightly modified value to that of \cite{Krueckel2015a} owing to an improved linear propagation loss estimate ($1.35\pm0.3$ dB/cm).
Importantly, two-photon absorption is absent at $1.55 \mic$. The waveguide dimensions are displayed in Fig. \ref{fig:SiN}(a).
The waveguide height was around 700 nm, with about $5\%$ variation across the wafer. This uncertainty translates into a range of group-velocity dispersion (GVD)  profiles bounded by the shadow areas in Fig. \ref{fig:SiN}(a). The dispersion is calculated using finite-element simulations (COMSOL) and includes both the dispersion of the material in bulk and the waveguide's geometry. The dispersion of the film is measured  up to $1.7 \mic$ and extrapolated up to $2.4 \mic$
using empirical relations from the bulk material dispersion.
The slight tilt in the waveguide's wall is a consequence of the etching process. The COMSOL calculations include this tilt, and showed that tilt variations have a negligible effect in the GVD compared to the uncertainty in the waveguide height or material dispersion.
For the TE mode, the resulting GVD has two zero-dispersion wavelengths (ZDWs) giving a broad region (spanning 700 nm) with flat and moderate anomalous dispersion, centered around 1550 nm, as  desired for efficient supercontinuum generation.

The experimental setup is shown in Fig. \ref{fig:SiN}(b). (Please see acronyms in the caption). An Er femtosecond fiber laser centered at 1555 nm produces a 90 MHz pulse train; each pulse has a 33 nm bandwidth (FWHM) supporting a transform-limited 105 fs (FWHM) Gaussian pulse. The pulses were coupled by an objective lens into a DCF to compensate for the accumulated chirp before the waveguide. NDFs were used to control the pulse energy, and an FPC controlled the polarization, followed by a piece of single-mode fiber spliced to a tapered lensed fiber (OZ Optics, based on SMF-28), which focused to a  $2.0\mic$ FWHM spot size. We used an intensity autocorrelator to estimate the pulse duration to 130 fs FWHM (assuming a Gaussian shape) at the waveguide entrance, revealing some remaining chirp in the pulse.
The maximum total power before the waveguide was 42 mW. The output from the waveguide was collected with a tapered lensed fiber. To record the supercontinuum, we used two OSAs with spectral ranges 600-1700 nm and 1200-2400 nm, whose spectra were overlapped to get the final spectrum. The coupling loss is estimated at $6.5\pm1.0$ dB/facet (TE) and $5.3\pm1.0$ dB/facet (TM), providing a maximum input pulse energy of 105 pJ (TE) and 140 pJ (TM) coupled into the waveguide.

\begin{figure}[tbp]
\centering
\includegraphics[width=\linewidth]{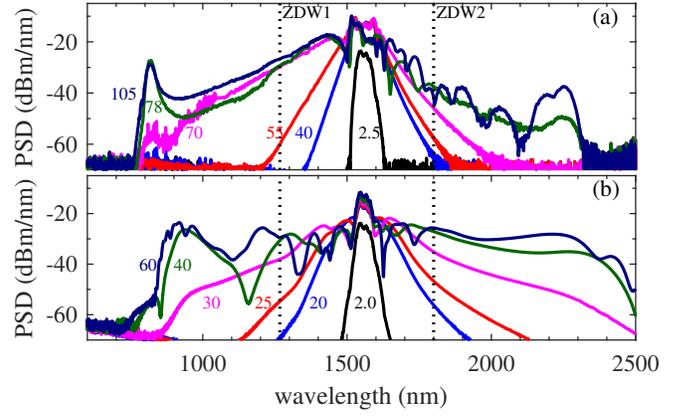}
\caption{(a) Experimental spectra in the TE case, showing the PSD at the end of the waveguide (corrected for end-facet coupling loss). The numbers show the estimated input pulse energies in pJ. (b) Results of numerical simulations of the TE FM using $h=660$ nm. The spectra show the total PSD at the waveguide end averaged over 50 noise realizations. The pulse was prechirped using ${\rm GDD}=+3500~{\rm fs^2}$.
}
\label{fig:TE-exp}
\end{figure}

The measured power spectral density (PSD) in the TE case is shown in Fig. \ref{fig:TE-exp}(a) for a range of pulse energies. The spectra for low pulse energies are clearly dominated by self-phase modulation (SPM), i.e., early stage broadening before soliton formation occurs. At 78 pJ, the soliton has formed, accompanied by two soliton-induced dispersive waves \cite{Skryabin:2010}, one on each side of the two ZDWs. The strongest dispersive wave is found at low wavelengths, peaked around 820 nm, while the mid-IR dispersive wave, peaked around 2250 nm, is less powerful and changes significantly when increasing the pulse energy further to the maximum value of 105 pJ (9 mW average power).

The results are verified with numerical simulations using the so-called nonlinear analytic envelope equation (NAEE) \cite{Conforti2013}, which resolves subcycle carrier-wave dynamics and includes a full expansion of the cubic nonlinearity. We only model the waveguide fundamental mode (FM).
The mode effective index and effective area from the COMSOL calculations were extrapolated directly to the NAEE grid without polynomial expansions and were carefully extended beyond the COMSOL modeling domain to avoid unphysical dispersion scenarios.
The frequency dependence of the effective mode area was modeled as shown in \cite{Laegsgaard:2007-OE}.
A constant propagation loss was used, and noise was included as a half photon per temporal mode in the input, modeling quantum vacuum fluctuations in the Wigner representation \cite{bache:2016-AIP}.
{\color{black}An extended NAEE model was used to include the Raman effect \cite{bache:2016-NAEE}, which empirically included the
Si-N asymmetric stretching mode centered at 410 $\rm cm^{-1}$ \cite{Ahmed2013}, having a broadband 70 $\rm cm^{-1}$ linewidth and a relative Raman strength $f_R=0.2$, both typical values for amorphous materials.}
Generally, the Raman effect gave minor contributions, in line with previous studies of silicon nitride waveguides. This might be due to the moderate anomalous dispersion range of the waveguide in which the Raman effect can influence the soliton by redshifting it.

\begin{figure}[tbp]
\centering
\includegraphics[width=1\linewidth]{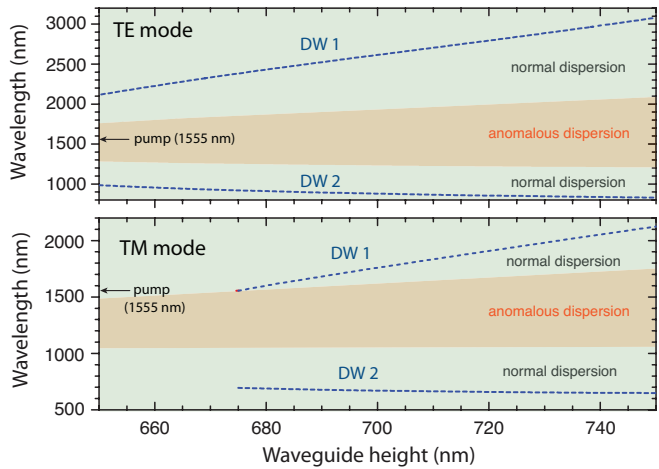}
\caption{Calculated ZDWs vs. waveguide height $h$ for TE and TM modes. The dispersive wave curves were calculated by the standard phase-matching formula $\beta_{\rm s}(\omega_{\rm DW})=\beta_{\rm lin}(\omega_{\rm DW})$ \cite{Skryabin:2010}
between a soliton at 1550 nm and a linear dispersive wave. Note that for the TM case $h>675$ nm to support a soliton at 1555 nm, so no DWs can be found for $h<675$ nm.}
\label{fig:ZDW}
\end{figure}

Figure \ref{fig:TE-exp}(b) shows corresponding numerical simulations where the major features of the experimental spectra are reproduced in the power sweep, except for the precise location of the dispersive waves. This is likely due to the uncertainty of the mid-IR dispersion and to the precise soliton wavelength (i.e., how much it is frequency-shifted by Raman or self-steepening effects). The dispersive waves also seem to be narrower in the experimental case. The simulations only model the FM, but from the higher-order modes (HOMs) given by the COMSOL simulations we estimate that around 55\% of the input energy will be coupled to the FM, and the rest goes into the HOMs. {\color{black}The HOMs will not show any spectral broadening due to their lower peak powers, higher GVD, and larger mode areas.} In the numerical spectra shown in Fig. \ref{fig:TE-exp}(b) we have therefore added the equivalent energy of the HOMs at the end of the simulation as 45\% of the total input spectrum. Additionally, combining the 33 nm bandwidth of the input pulse with the measured value of 130 fs FWHM corresponds to a group-delay dispersion of ${\rm GDD}=\pm 3500~{\rm fs}^2$. The simulations used positive chirp but similar results were found with negative chirp.
We note that the simulations used lower energies than the experiments, which can be attributed to uncertainties in coupling efficiencies, waveguide nonlinearity, dispersion, input pulse pre-chirp etc.

Figure \ref{fig:ZDW} shows that the waveguide height is a powerful dispersion tuning parameter. At the pump wavelength, the TE mode has anomalous dispersion for all considered heights,
while the GVD for the TM mode changes sign at $h=675$ nm. {\color{black}The simulations} show the best match to the experimental results for $h=660$ nm. With this parameter, the long- and short-wavelength dispersive waves (DW 1 and 2) agree quite well with the experiments.
This height is within the 5\% variation margin across the wafer as mentioned above. Similar results were observed for $h=$650$-$675 nm, while, beyond this range, the simulations did not match the experiments very well.

\begin{figure}[tbp]
\centering
\includegraphics[width=\linewidth]{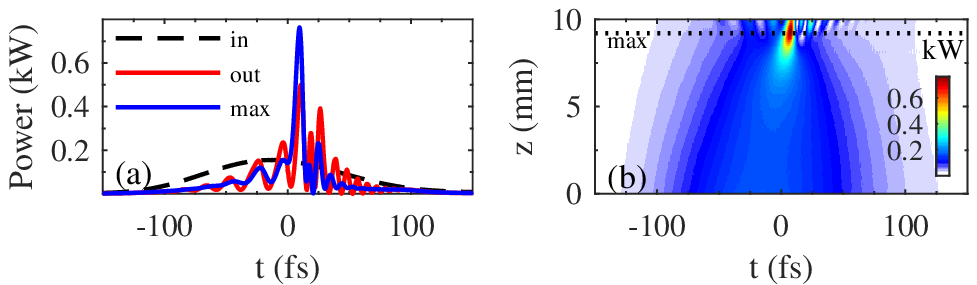}
\includegraphics[width=\linewidth]{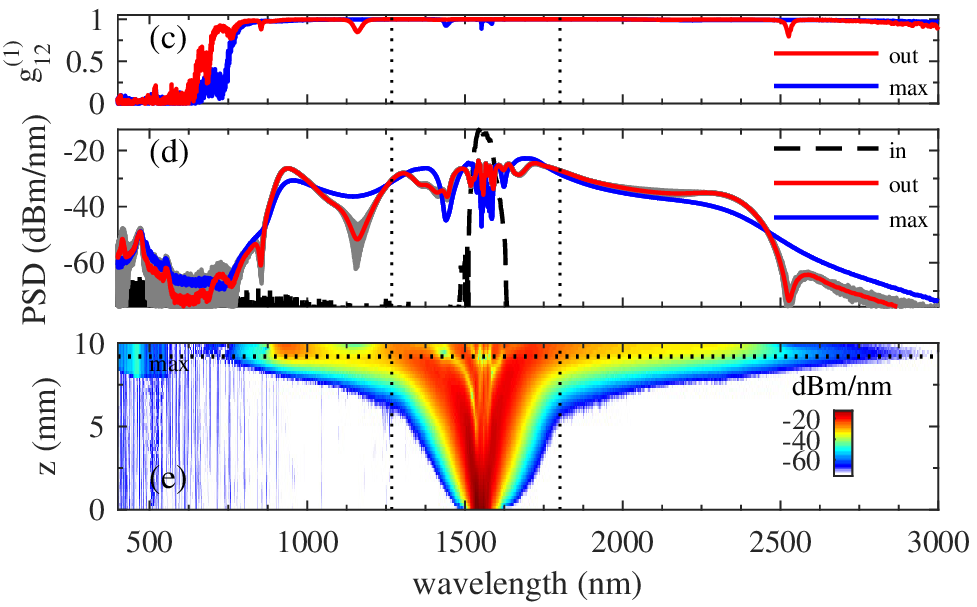}
\caption{Numerical simulation of the TE-polarization FM, using 40 pJ total input energy, as in Fig. \ref{fig:TE-exp}(b). (a) and (b) show time domain and (e) shows wavelength domain dynamics of a single noise realization. The  coherence function (c) and FM spectra (d) were averaged over 50 noise realizations (each shown in gray for the output pulse).}
\label{fig:sim-TE}
\end{figure}

Figure \ref{fig:sim-TE} shows a representative simulation for 
{\color{black}the supercontinuum propagation dynamics.} 
The pulse energy is 40 pJ, corresponding to the green curve in Fig. \ref{fig:TE-exp}(b).
In Fig. \ref{fig:sim-TE} (a) the soliton forms as a single-cycle spike (5 fs FWHM), and, after this self-compression point ("max"), the temporal trace shows interference oscillations. These are due to the two dispersive waves phase-matched to the soliton, all temporally overlapping, but located at different wavelengths. In the spectral evolution (e) the dispersive waves are seen to be present at the soliton formation point ("max"), and, after this, they grow significantly; see also (d).
In (c), the shot-to-shot coherence is shown, as calculated from the first order coherence function $g_{12}^{(1)}(\lambda)$ \cite{Dudley2002}. This excellent coherence
pertains for lower powers, while we found it degrades for higher levels (which is quite typical).

\begin{figure}[tbp]
\centering
\includegraphics[width=\linewidth]{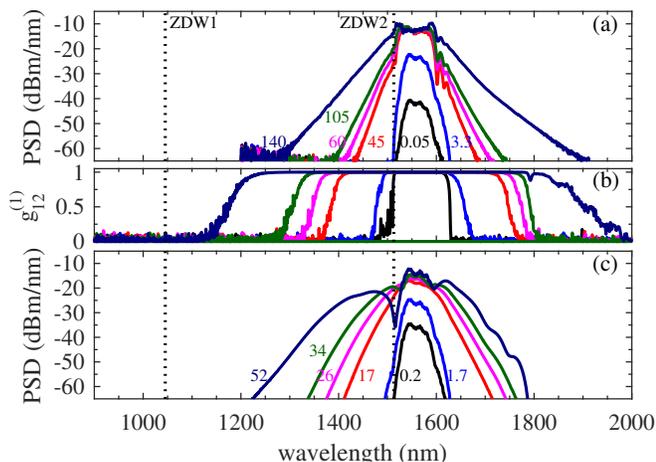}
\caption{(a) Experimental spectra for various TM pump pulse energies; The ZDWs are those calculated for {\color{black}the TM mode} using a 660 nm waveguide height. TM mode numerical simulations showing (b) the coherence function and (c) the average spectra.
The simulations used the same waveguide specs (width, height) and input pulse prechirp as Fig. \ref{fig:TE-exp}(b).}
\label{fig:TM-exp}
\end{figure}

Figure \ref{fig:TM-exp}(a) shows the TM case. While {\color{black}less} broadening was observed, the spectrum broadens sufficiently to enter the important 1600-1850 nm wavelength range for 3-photon absorption microscopy \cite{Horton2013}.
The spectral shapes are indicative of optical wave breaking \cite{Tomlinson:1985}, which happens when the pump dispersion is normal, so no soliton can form.
The spectral broadening is typically much weaker than in the soliton case because no dispersive waves are formed and the pulse does not self compress.
Instead, the pulse becomes highly chirped and eventually develops steep temporal shock fronts.
In frequency domain, this is accompanied by highly coherent SPM-induced spectral broadening.
To understand when TM has normal GVD at 1555 nm, we refer to Fig. \ref{fig:ZDW}, which shows that the TM mode has normal GVD at 1555 nm when
the waveguide height is taken well below 700 nm.
Indeed, the numerical simulations shown in Fig. \ref{fig:TM-exp}(c) agree well with the experimental results using the exact same parameters as in the TE case concerning {\color{black}waveguide size} 
and pulse pre-chirp.
In time domain (not shown), the pulses were chirped, but the powers were too low to observe severe shock front formation.
{\color{black}The experimental data show extremely symmetrical SPM broadening, which could only be observed in the simulations} 
when using positively chirped input pulse, while a negative chirp tended to give more asymmetric spectra. Finally, Fig. \ref{fig:TM-exp}(b) shows the coherence of the simulated TM continua, demonstrating perfect coherence across the generated bandwidth, as is
typical for optical wave breaking.

\begin{figure}[tbp]
\centering
\includegraphics[width=0.97\linewidth]{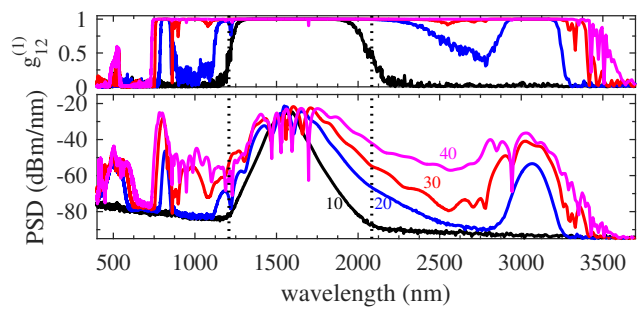}
\caption{Simulations showing the spectral coherence and average PSD  for a 750 nm waveguide height and using a transform-limited TE polarized pump pulse.}
\label{fig:TE-750}
\end{figure}

Our simulations show that for a transform-limited pump, the results in Fig. \ref{fig:TE-exp}(a) could be obtained using substantially lower pump energies.
Additionally, the long pump wavelength of Er-fiber lasers allows generating a mid-IR dispersive wave in the TE case for thick waveguides (Fig. \ref{fig:ZDW}). Combining these efforts, Fig. \ref{fig:TE-750} shows the promising potential of these waveguides: with a $h=750$ nm thick waveguide and a TE transform-limited pump {\color{black}with only 20 pJ energy, over 2 octaves of coherent supercontinua are generated, extending well into the mid-IR.}

Summarizing, we studied supercontinuum generation in a silicon-rich nitride waveguide pumped with low-energy femtosecond pulses (80-140 pJ) from an Er fiber laser.
In the TE case, a soliton and two dispersive waves were excited to give a 1.5 octave supercontinuum (820-2250 nm at $-30$ dB). In the TM case, a continuum was generated by optical wave breaking as the TM mode had normal dispersion at the pump wavelength. Numerical results indicate an excellent coherence of the supercontinua and that they could extend into the mid-IR with a slightly thicker waveguide.
These results are promising for short-range near-IR and mid-IR coherent supercontinuum generation in an easily manufacturable CMOS-compatible waveguide.

\textbf{Funding.} {\color{black}Vetenskapsr\aa det;} Teknologi og Produktion, Det Frie Forskningsr\aa d (FTP, DFF) (11-106702); European Research Council (ERC) (ERC-AdvG 291618).

\end{document}